\documentclass[letter]{aa} 
\usepackage{graphicx}
\usepackage{txfonts}
\newcommand{\hi}{{\rm H{\textsc{i}}\,}}
\newcommand{\hii}{{\rm H{\textsc{ii}}\,}}

\newcommand{\hei}{{\rm He{\textsc{i}}\,}}
\newcommand{\heii}{{\rm He{\textsc{ii}}\,}}

\newcommand{\msolar}{\mbox{\,M$_{\odot}$}}

\newcommand{\ovi}{{\rm O{\textsc{vi}}\,}}

\newcommand{\novi}{{\rm N(O{\textsc{vi}})\,}}

\newcommand{\civ}{{\rm C{\textsc{iv}}}}
\newcommand{\nciv}{{\rm N(C{\textsc{iv}})\,}}
\newcommand{\nnv}{{\rm N(N{\textsc{v}})\,}}

\newcommand{\nv}{{\rm N{\textsc{v}}\,}}

\begin{document}
\title{A non-equilibrium ionization model of the Local Bubble (I)}
\subtitle{Tracing \civ, \nv, and \ovi ions}

   \author{Miguel A. de Avillez\inst{1} \and Dieter Breitschwerdt\inst{2}}
   \institute{Department of Mathematics, University of \'Evora, R. Romao Ramalho 59, 7000 \'Evora, Portugal\\
   \email{mavillez@galaxy.lca.uevora.pt}
   \and
    Zentrum f\"ur Astronomie und Astrophysik, Technische Universit\"at Berlin, Hardenbergstrasse 36, D-10623 Berlin,
Germany \\
    \email{breitschwerdt@astro.physik.tu-berlin.de}
    }

    \date{Received May 2, 2011; accepted Month Day, 2011}

   \titlerunning{A NEI Model of the Local Bubble (I) - Tracing \civ, \nv and \ovi Ions}
   \authorrunning{M.A. de Avillez and D. Breitschwerdt}


  \abstract
  {}
  {We present the first high-resolution non-equilibrium ionization simulation of the joint evolution of the
Local Bubble (LB) and Loop I superbubbles in the turbulent supernova-driven interstellar medium (ISM). The
time variation and spatial distribution of the Li-like ions \civ, \nv, and \ovi inside the LB are studied in
detail.}
  {This work uses the parallel adaptive mesh refinement code EAF-PAMR coupled to the
newly developed atomic and molecular plasma emission module E(A+M)PEC, featuring the time-dependent
calculation of the ionization structure of H through Fe, using the latest revision of solar abundances. The finest AMR
resolution is 1 pc within a grid that covers a representative patch of the Galactic disk (with an area of 1 kpc$^{2}$ in
the midplane) and halo (extending up to 10 kpc above and below the midplane).}
{The evolution age of the LB is derived by the match between the simulated and observed absorption features
of the Li-like ions \civ, \nv, and \ovi. The modeled LB current evolution time is bracketed between 0.5 and
0.8 Myr since the last supernova reheated the cavity in order to have $\novi<8\times 10^{12}$ cm$^{-2}$,
$\log[\nciv/\novi]<-0.9$ and $\log[\nnv/\novi]<-1$ inside the simulated LB cavity, as found in \textsc{Copernicus},
IUE, GHRS-IST and FUSE observations.}
{}
   \maketitle
%

\section{Introduction}

The ionization structure of the Local Bubble (LB), like that of any superbubble, should in principle reflect
its evolutionary state for given boundary (and initial) conditions, if we assume a multi-supernova/stellar
wind origin (e.g, Innes \& Hartquist 1984; Cox \& Anderson 1982; Smith \& Cox 2001). While the ambient medium
at the time of the first supernova (SN) explosion is unknown, the subsequent time-dependent energy input rate is
-- quite surprisingly -- much more tightly constrained, although no early-type stars (or relics thereof)
have been found within the current LB volume. The kinematic analysis of nearby stars (Ma\'{\i}z-Apell\'aniz 2001),
moving groups (Bergh\"ofer \& Breitschwerdt 2002; BB02), and the whole volume of 400 pc in
diameter centered on the Sun from Hipparcos data (Fuchs et al. 2006) consistently point to $14-19$ stars having
exploded during the past 15 Myr on their way through the region now occupied by the LB.

Collisional ionization equilibrium (CIE) simulations of the LB powered by the missing stars of the subgroup B1
of Pleiades (BB02) within a turbulent interstellar medium (Breitschwerdt \& de Avillez 2006 (BA06); de Avillez \&
Breitschwerdt 2009 (AB09)) reproduced the clumpy distribution of \ovi, as well as its column density and dispersion
inside the LB cavity observed by Copernicus (Jenkins 1978; Shelton \& Cox 1994) and FUSE (Oegerle et al. 2005; Savage \&
Lehner 2006), and the lack of \civ\ and \nv ions in the bubble, which is consistent with the observations with
Copernicus (York 1977), IUE (Bruhweiler et al. 1980), and GHRS-HST (Bertin et al. 1995; Huang et al. 1995). These
observations point to column densities $\nciv<7\times 10^{11}$, $\nnv <3\times 10^{12}$
cm$^{-2}$, $\log [\nciv/\novi]\lesssim-0.9$ and $\log [\nnv/\novi]\lesssim-1.0$. In that CIE model, the
match between the simulated and observed column densities indicated that the last SN event occurred 0.5 Myr ago.

However, significant deviations from global CIE conditions are likely to occur (e.g., Kafatos 1973) as suggested
by observations of the LB in the EUV (Jelinsky et al. 1995; Hurwitz et al. 2005) and X-ray (Sanders et al. 2001;
McCammon et al. 2002; Freyberg \& Breitschwerdt 2003; Henley et al. 2007) bands and by non-equilibrium ionization (NEI)
one-dimensional simulations of the LB (e.g., Cox \& Anderson, 1982; Smith \& Cox 2001).

The present work explores the degree to which non-equilibrium ionization (NEI) affects the conclusions of
otherwise similar CIE modeling. We extend our previous CIE work (BA06 and AB09) to (i) model the LB evolution
hydrodynamically and trace its ionization structure (and that of the surrounding ISM) in a time-dependent fashion, (ii)
study the spatial distribution of the Li-like ions \civ, \nv, and \ovi and compare them with observations, and
(iii) constrain the LB age by matching simulated and observed line-of-sight distributions of these ions.
This letter is organized as follows: Section 2 summarizes the new model setup and simulation. In Section 3, the
distribution of Li-like ions and their column density ratios within the LB are discussed and compared to
observations. The discussion of the results (Section 4) closes the paper.

We simulate the effects of the explosions of the stars of a moving subgroup, the living members of which now belong to
the UCL and LCC associations, as their trajectories have crossed the LB volume towards their present position.

\section{Model and numerical setup}

We simulate the simultaneous evolution of the Local and Loop~I superbubbles. These are the result
of the successive explosions of massive stars from, respectively, a moving subgroup consisting of 17 stars
with masses $\in[8.2,21.5]$ \msolar, whose living members now belong to the UCL and LCC associations, as their
trajectories have crossed the LB volume towards
their present position (Fuchs et al. 2006), and Sco Cen consisting of 39 stars with masses $\in[14,31]$ \msolar\ (Egger
1998). Their time of explosion is calculated by the Fuchs et al. (2006) formula. The simulation setup is similar to that
discussed in BA06 and AB09, but with three major differences from these previous CIE models: Firstly, the
\emph{Time-dependent evolution} of the ionization structure of H, He, C, N, O, Ne, Mg, Si, S, and Fe ions with Asplund
et al. (2009) abundances derived using the Eborae Atomic+Molecular Plasma Emission Code (E(A+M)PEC\footnote{The code is
written in OpenCL and runs in AMD and NVIDIA Graphics Processor Units (GPUs). A description of the software and
ionization fractions, cooling and emission spectra tables can be found at http://www.lca.uevora.pt/research.html See
also references therein.}; Avillez et al. 2011). The physical processes included in E(A+M)PEC are electron impact
ionization, inner-shell excitation auto-ionization, radiative and dielectronic recombination, charge-exchange reactions
(recombination with \hi and \hei and ionization with \hii and \heii), continuum (bremsstrahlung, free-bound, two-photon)
and line (permitted, semi-forbidden and forbidden) emission in the range 1\AA~-610 $\mu$, and molecular lines.
The E(A+M)PEC code also includes inner shell photoionization, ionization of \hi by Lyman continuum photons
produced by the recombination of helium, and molecular chemistry involving H, C, and O. The cooling function
(calculated at each cell of the grid at all
time steps) includes these processes, while the spectra calculations include line and continuum emissions. The
second difference is that local self-gravity and heat-conduction (Balbus \& Mckee 1982; Dalton \& Balbus 1993) are
calculated at every time step; Finally, the coarse grid resolution is 8 pc, while the finest AMR resolution is 1
pc - the highest so far used for LB evolution calculations. Periodic and outflow boundary conditions are set along the
vertical faces and both the top and bottom ($z=\pm 10$ kpc) of the grid, respectively.

\begin{figure}[thbp]
\centering
\includegraphics[width=0.72\hsize,angle=-90]{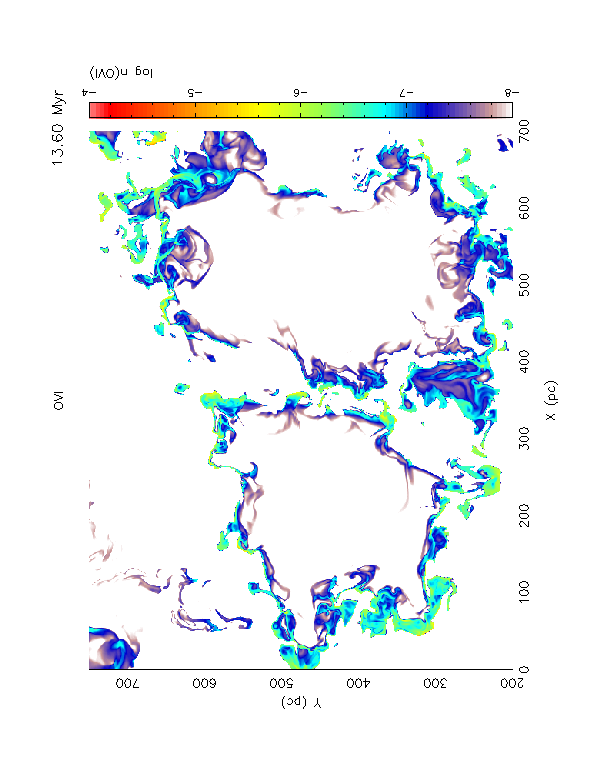}\\
\includegraphics[width=0.72\hsize,angle=-90]{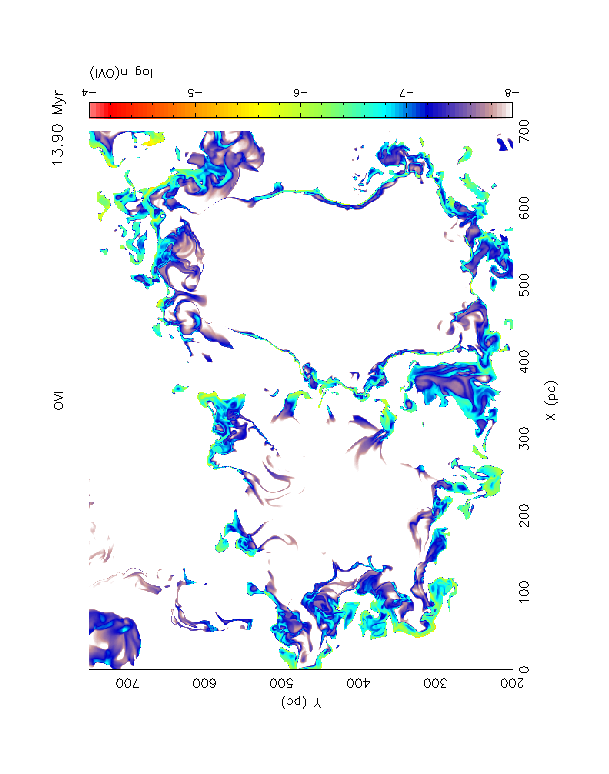}
\caption{\ovi density distributions (cuts through the Galactic midplane) in the LB centered at
($x=200, y=450$) pc and Loop I (the adjacent bubble on the right) at 0.5 (top) and 0.8 (bottom) Myr after the last SN
in the LB, which occurred at evolution time 13.1 Myr. Both bubbles are surrounded with a thin fragmented \ovi
bearing shell.}
\label{ionsevol}
\end{figure}

\begin{figure}[thbp]
\centering
\includegraphics[width=0.98\hsize,angle=0]{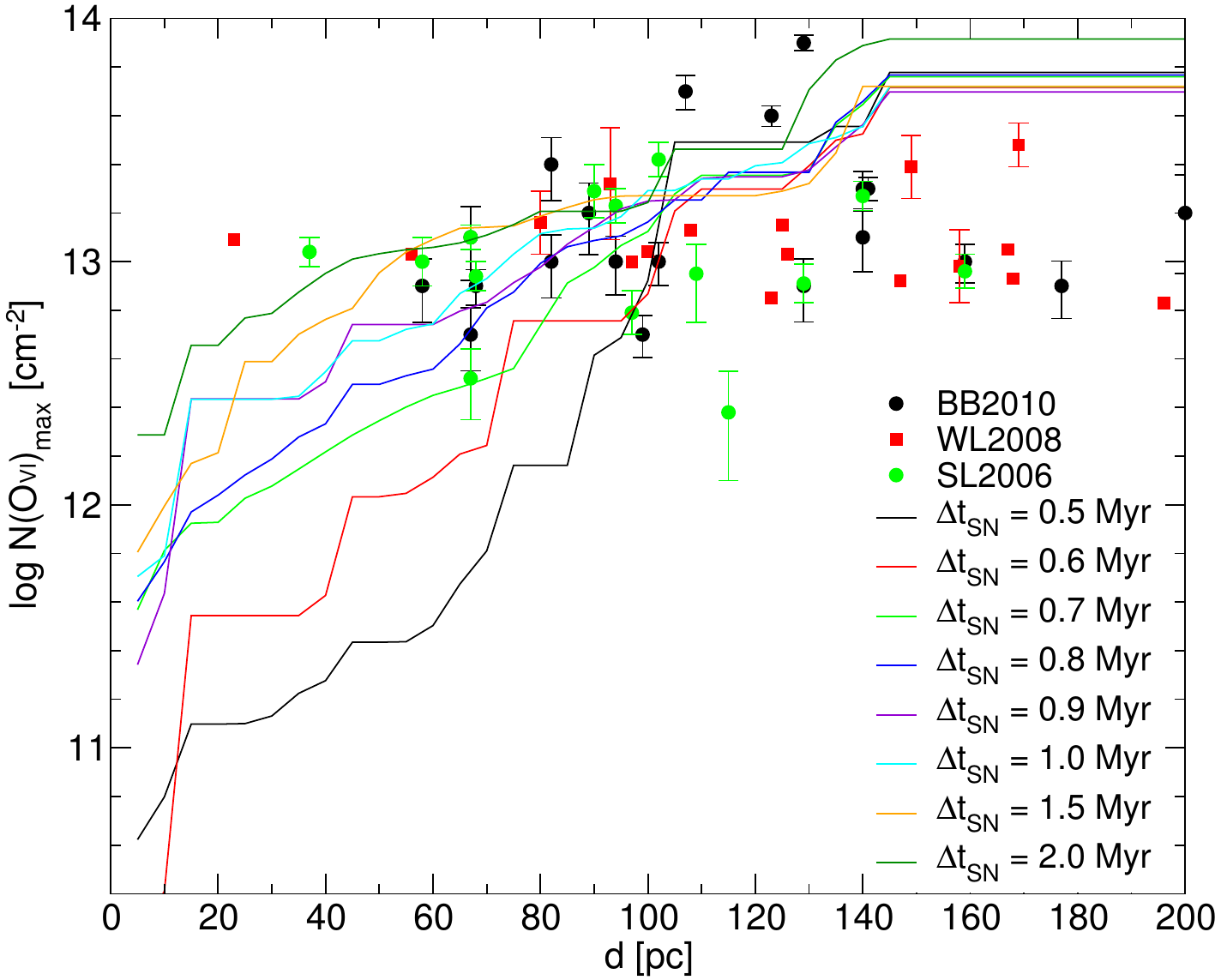}
\caption{Maximum (top) model \ovi column densities in a 200 pc radius sphere centered at the Sun at times 0.5 through
1.0 Myr after the last SN event in the LB, overlaid by FUSE data from Savage \& Lehner (2006; SL2006), Welsh \&
Lallement (2008; WL2008) and Barstow et al. (2010; BB2010) for observations agains white dwarfs in the up to 200 pc
from the Sun. The horizontal axis indicates the integration distance for the model and the star distance from the
observational data.}
\label{oviaverage}
\end{figure}

\section{Results}

To model the Local Bubble, which is assumed to be embedded in a realistic background ISM, the simulations were run
for a considerable time in order to establish the full Galactic fountain and a global dynamical equilibrium
with the density, pressure, and temperature distributions conditions similar to those described in de Avillez \&
Breitschwerdt 2004). At an ISM evolution time of 250 Myr, a tracker searches for a molecular region of sufficient mass
to form the ScoCen stars (adopting a 5\% star formation efficiency). The UCL and LCC clusters paths towards ScoCen
are then set using the kinematic parameters derived by Fuchs et al. (2006) with the last SN event in the LB
cavity occurring 13.1 Myr after the first explosion (that is, after an ISM evolution time of 263.1 Myr). The
Local and Loop I bubbles evolve in time interacting for most of the LB lifetime, and eventually the two bubbles
merge with Loop I material expanding into the LB cavity. Here, we discuss the evolution of the LB in the first
Myr after the last SN explosion.

\subsection{\ovi distribution in the LB}

Figure~\ref{ionsevol} displays time snapshots of the\, \ovi density distribution within the LB at an evolution time
(since the first SN in the LB) of t$_{evol}=13.6$, and 13.9 Myr, that is $\Delta \tau \in \{0.5, 0.8 \}$ Myr,
respectively, after the last SN event within the LB cavity. The \ovi has a clumpy distribution resulting from
local cooling and turbulent motions, driven by shear flows inside the LB cavity, whose highest values occur
towards the boundary.

\begin{figure}[tbhp]
\centering
\includegraphics[width=0.95\hsize,angle=0]{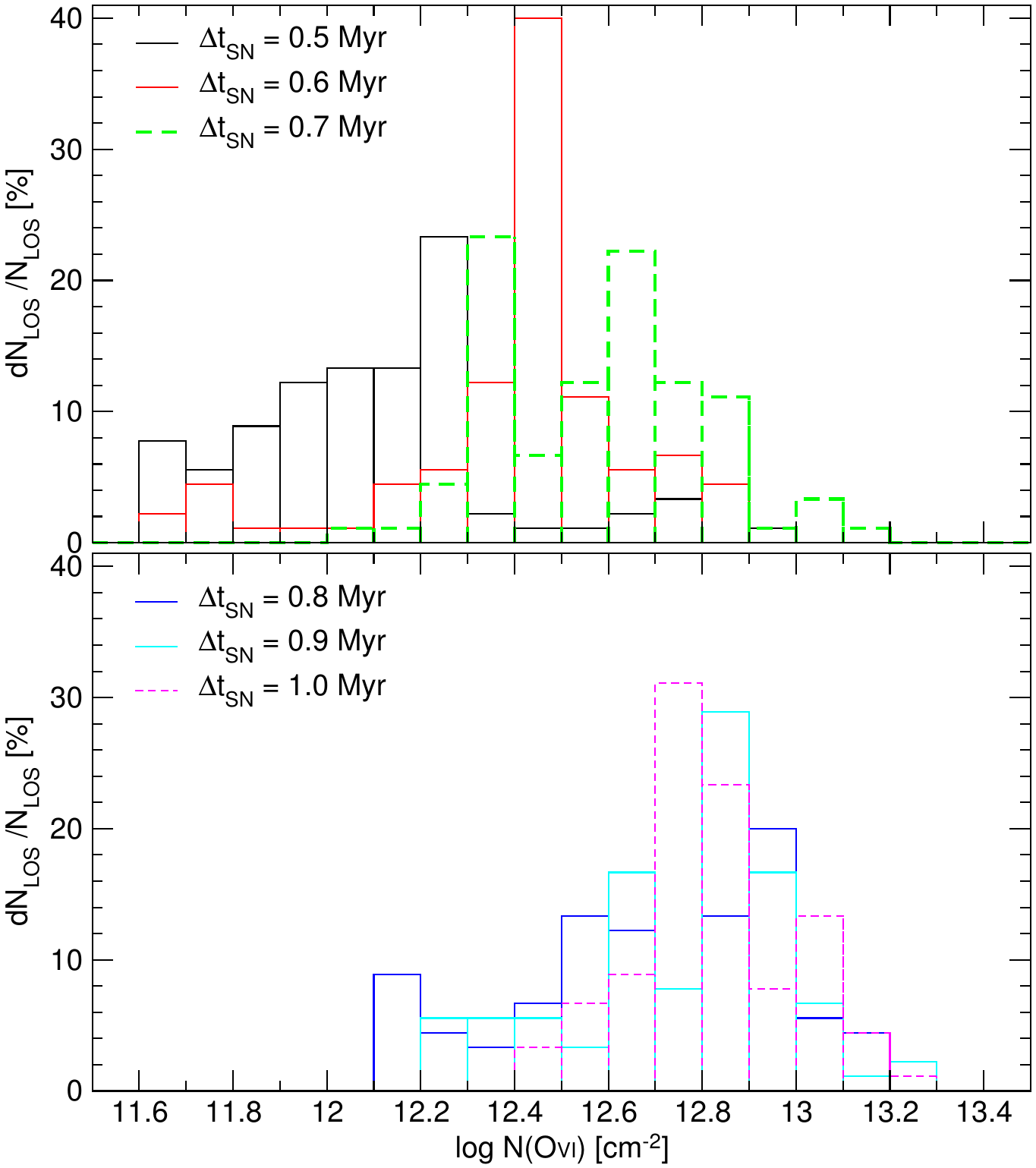}
\caption{Histograms of the \novi model predictions for LOS sampling only gas inside the cavity, i.e., LOS
lengths $< 100$ pc, between 0.5 and 1.0 Myr after the last explosion in the LB cavity.}
\label{ovihist}
\end{figure}

The spatial and temporal variations in \ovi are explored by calculating the \ovi column density (\novi) through
LOS taken from the Sun at $(x=200, y=450)$ pc, 100 pc from the interaction region between the Local and Loop~I
bubbles. The column densities were calculated with a 5 pc step length out to a maximum of 200 pc, spaced at 1 degree
intervals to examine the full 200 pc radius sphere about the assumed Solar position. Figure~\ref{oviaverage}
displays the maximum \novi values over all the lines of sight at times 0.5 Myr through to 1, 1.5 and 2.0 Myr
after the last SN explosion in the cavity. In addition, the figure also shows FUSE data against white dwarfs
(Savage \& Lehner 2006 (SL2006); Welsh \& Lallement 2008 (WL2008); Barstow et al. 2010 (BB2010)). 

The maximum \novi increases with both path length and time, being lower than $10^{13}$ cm$^{-2}$ inside the
LB up to a radius of 100 pc in the first 0.5-0.6 Myr since the last SN and growing to a maximum of $\log \novi=13.3$ (in
units of cm$^{-2}$) in the next 0.2 Myr. For distances $>140$ pc, the lines of sight are sampling gas not only from Loop
I but also from the surrounding interstellar medium, hence the increase in the column density. It is clear that the
modelled column densities fall within and follow the observed (with FUSE) data for path lengths greater than 50 pc. With
increasing time, there is a slight increase in the amount of \ovi at smaller distances ($d<50$ pc; orange and dark green
lines at times 1.5 Myr and 2.0 Myr after the last SN event in the cavity).

For the first million years of evolution of the simulated bubbles, the majority of the LOS ($>75\%$) samples gas with
$\log$(\novi)$<13$, with this number reaching 99\% and 95\% levels at 0.5 Myr and 0.7 Myr after the last explosion (top
panel of Figure~\ref{ovihist}) and decreasing to some 80\% in the next 0.3 Myr (bottom panel of Figure~\ref{ovihist}).

\begin{figure*}[thbp]
\centering
\includegraphics[width=0.47\hsize,angle=0]{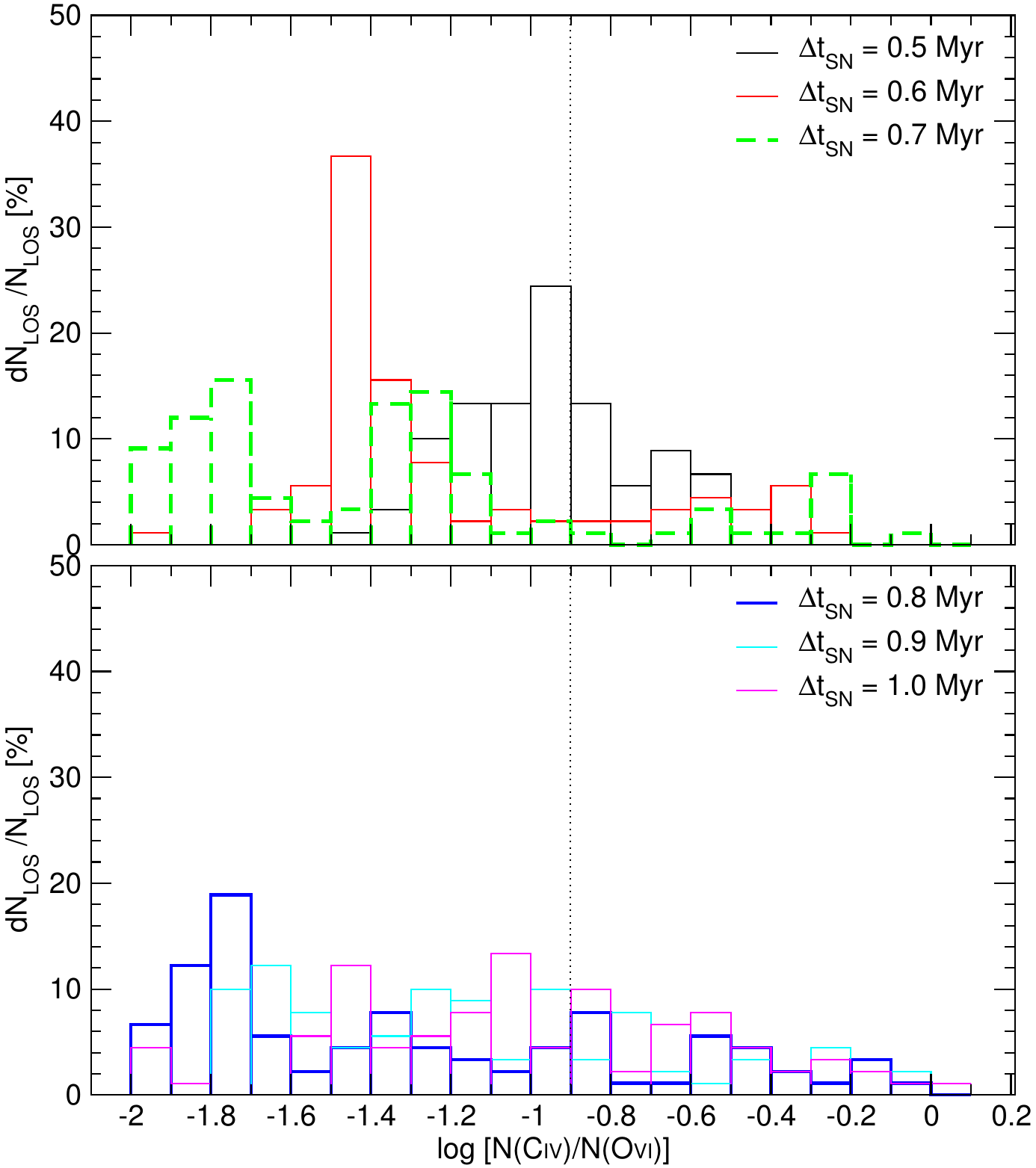}
\includegraphics[width=0.47\hsize,angle=0]{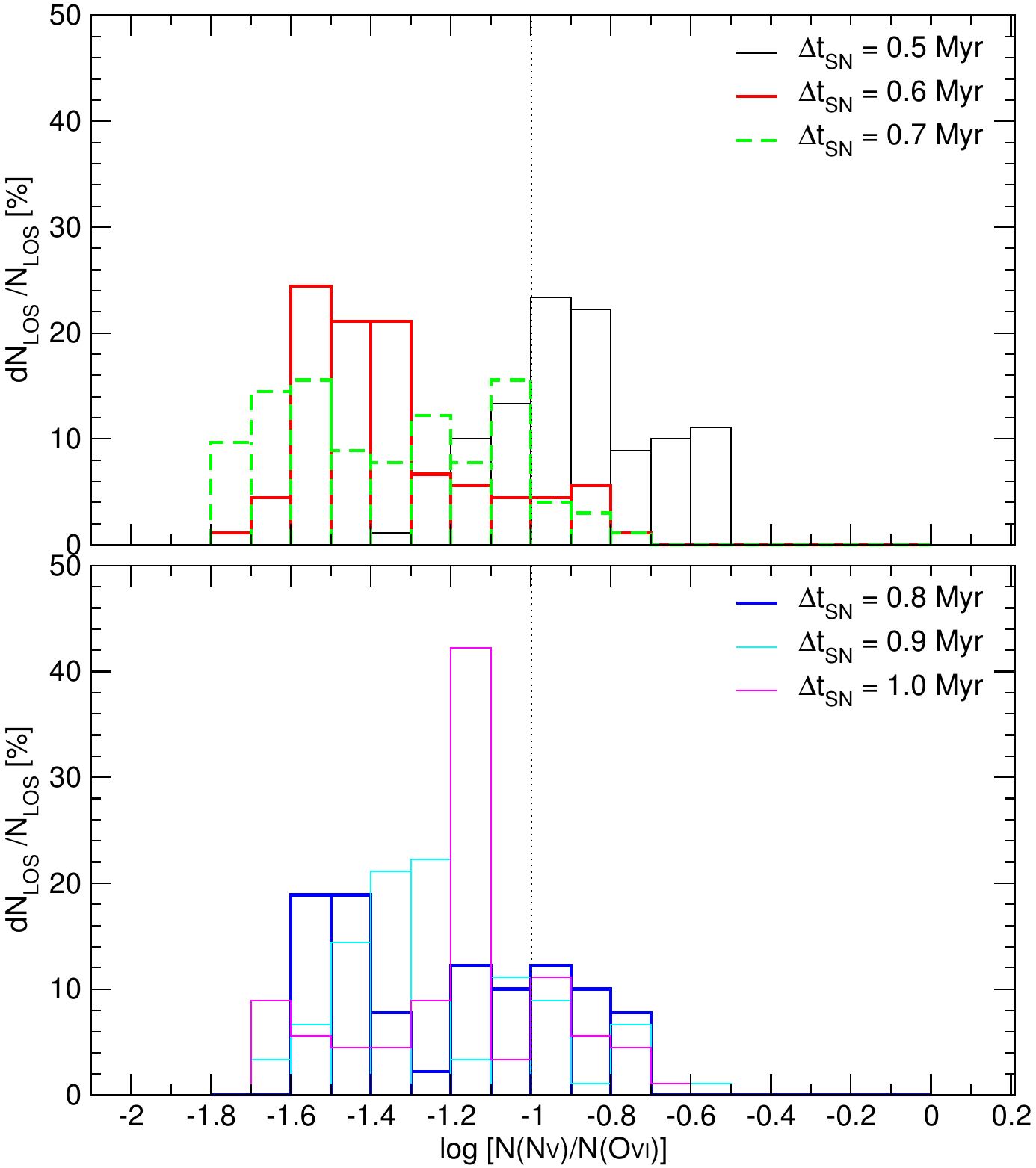}
\caption{Histogram of the LOS measurements of the column density ratios \nciv/\novi (left panel)
and \nnv/\novi (right panel) inside the simulated LB ($l_{LOS}<$ 100 pc) for 0.5-1.0 Myr after the last SN
explosion. The vertical dotted line marks the critical column density ratios of $-0.9$ (left) and $-1.0$ (right),
respectively (for details see text).}
\label{histratios}
\end{figure*}

\subsection{Distribution of Li-like ion ratios}

Figure~\ref{histratios} displays the histograms of the LOS ratios of $\log[\nciv/\novi]$ and $\log[\nnv/\novi]$ in the
LB between 0.5 Myr and 1.0 Myr after the last SN event. The fraction of LOS sampling gas with column density ratio
$\log[\nciv/\novi]<-0.9$ increases with time towards 87\% at 0.6 Myr and 93\% at 0.7 Myr (top panel, left column of
Figure~\ref{histratios}) then decreases to less than 80\% in the next 0.1 Myr of the LB evolution (bottom panel, left
column of the figure). The small fraction of LOS with $\log[\nciv/\novi]> -0.9$ is consistent with the detection
by Welsh et al. (2010) of interstellar \civ~in the LB with $\nciv\simeq 10^{12}$ cm$^{-2}$ towards a star located at 74
pc from the Sun. For $\log[\nnv/\novi]<-1.0$, the fraction of LOS sampling gas up to this threshold also increases to
89\% and 92\% levels at 0.6 and 0.7 Myr, respectively, after the last explosion in the LB, followed by a steep
decrease to less than 70\% in the next few hundred thousand years (right column of Figure~\ref{histratios}). This
behaviour of the column density ratios in the simulated LB confirms the presence of a very small amount of
\civ~and \nv ions and their uncorrelated distribution with \ovi.

\section{Discussion and conclusions}

The work described here focuses on the detailed numerical evolution of a multi-supernova origin of the LB,
following the studies by BA06 and AB09. A major difference from previous studies by other authors is that we
(i) follow in a Eulerian fashion the joint evolution of the Local and Loop I bubbles in the turbulent ISM, (ii) use a
three-dimensional high resolution supernova-driven ISM model to simulate a realistic ambient medium, (iii) calculate the
\emph{time-dependent evolution of the complete ionization structure}, and therefore also the corresponding
\textit{time-dependent cooling function} at each cell of the computational grid, and (iv) include the latest
revision of solar abundances. 

Three conditions determine the present evolution time ($t_{evol}$) of the LB: the $\novi$ threshold of
$8\times 10^{12}$ cm$^{-2}$ and the ratios $\log[n(\civ)/n(\ovi)<-0.9$ and $\log[\nnv/\novi]<-1.0$. A
comparison between the LOS measurements at different times in the simulated LB indicates that $t_{evol}$ can be
constrained between 0.5 Myr and 0.8 Myr since the last SN explosion. The other times are excluded because they
do not satisfy simultaneously the three threshold conditions, therefore, contradicting \textsc{Copernicus},
IUE, and GHRS-HST observations. It should be kept in mind, however, that the age of the LB, solely based on
the simulated \ovi, is between 0.5 Myr and 1.1 Myr since the last SN event. The derived LB present age
corresponds to a delay of at most 0.3 Myr (0.6 Myr if we only consider the \ovi distribution) compared to the
0.5 Myr obtained in the CIE simulations of BA06 and AB09. 

The comparison of Li-like ions between observations and NEI simulations pins down for the first time the age of the LB
reliably well, which was impossible in previous CIE simulations. In particular, modelling the densities
of \emph{different} ions with different ionization and recombination timescales gives \emph{abundance-independent}
results. Moreover, since spectral resolution in our models is not an issue, we will be able in the near future to
also compare theoretical line widths to observations.

The small column densities observed over a wide range of temperatures in the simulations are the result of delayed
recombination, which also manifests itself in X-ray emission at low temperatures, as we will discuss
in the next few papers of the series. Finally, we note that we have performed calculations of the distribution of
$^{60}{\rm Fe}$ in the solar neighbourhood and compared the sequence of explosions with the measured iron peak in the
deep sea ferromanganese crust (Knie et al. 2004), including the new half lifetime of Rugel et al. (2009), and found
excellent agreement in the time sequences (Breitschwerdt et al. 2012). Further developments of the model comprising
the inclusion of magnetic field and cosmic rays will also be explored in papers to follow.

\begin{acknowledgements}
We would like to thank the referee, Don Cox, for his detailed reports, constructive criticism, and valuable
suggestions that allowed us to improve this letter significantly. The simulations were carried out at the Milipeia
Supercomputer (Center for Computational Physics, University of Coimbra) and at the ISM-Cluster of the
Computational Astrophysics Group (Dept. of Mathematics, University of \'Evora). This research is supported by the
FCT (Portugal) project PTDC/CTE-AST/70877/2006.

\end{acknowledgements}

\end{document}